\documentstyle[aps,prl,psfig]{revtex}
\begin{document}

\title{Local optical spectroscopy of semiconductor nanostructures
in the linear regime}

\author{Oskar Mauritz, Guido Goldoni, and Elisa Molinari}
\address{Istituto Nazionale per la Fisica della Materia (INFM),\\ 
and Dipartimento di Fisica,
Universit{\`a} di Modena e Reggio Emilia, 
Via Campi 213A, I-41100 Modena, Italy}
\author{Fausto Rossi}
\address{ Istituto Nazionale per la Fisica della Materia (INFM),\\
and Dipartimento di Fisica,
Politecnico di Torino \\
Corso Duca degli Abruzzi 24, I-10129 Torino, Italy}

\date{\today}

\maketitle

\begin{abstract}

We present a theoretical approach to calculate the {\it local}
absorption spectrum of excitons confined in a semiconductor
nanostructure. Using the density-matrix formalism, we derive a
microscopic expression for the non-local susceptibility, both in the
linear and non-linear regimes, which includes a 
three-dimensional description of electronic quantum states and their
Coulomb interaction. The knowledge of the non-local susceptibility
allows us to calculate a properly defined local absorbed power, that
depends on the electro-magnetic field distribution.

We report on explicit calculations of the local linear response
of excitons confined in single and
coupled T-shaped quantum wires with realistic geometry and
composition. We show that significant interference effects in the 
interacting electron-hole wavefunction induce 
new features in the space-resolved optical spectra, particularly 
in coupled nanostructures. When the spatial extension of the 
electromagnetic field is comparable to the exciton Bohr radius,
Coulomb effects on the local spectra must be taken into account for a 
correct assignment of the observed features. 

\end{abstract}


\section*{Introduction}


The recent achievements in the field of semiconductor nanostructures
have prompted a strong effort in developing local experimental probes,
in order to obtain spatial maps of the nanostructures and their
quantum states. While conventional optical spectroscopy gives
information on a large region containing thousands of nanostructures,
confocal diffraction-limited microscopy has allowed the investigation
of individual nanostructures \cite{confocal}. To probe the spatial
distribution of quantum states, the spatial resolution must be reduced
much below the optical wavelength; this has been obtained by means of
near-field scanning optical microscopy (NSOM) \cite{SNOM}. In semiconductor
quantum wires \cite{WIRES} and dots \cite{DOTS} the resolution of these
experiments has been increasing in recent years.

From the theoretical point of view it was soon recognized that the
interpretation of NSOM spectroscopic data requires to take into
account the effects of the fiber tip and dielectric discontinuities on
the electromagnetic (EM) field generated in the sample. For example,
the near-field distribution of the EM-field \cite{chang97} and its
interaction with arrays of point-like particles \cite{hanewinkel97}
have been studied in detail.

On the other hand, the interactions of a highly inhomogeneous EM-field
with the quantum states in the semiconductor nanostructures received
much less attention \cite{bryant98}. A theoretical effort in this
direction is important for different reasons. First, when the dipole
approximation is abandoned and the non-local response of the medium is
taken into account, local absorption itself is in principle ill
defined (i.e., it is not independent of the EM-field distribution, as
we will show); a general theoretical reformulation is therefore
required. In addition, it may be expected that spatial interference of
quantum states plays an important role when variations of the
electromagnetic field occur on a ultra-short length scale, i.e., on
the scale of the Bohr radius; hence, the necessity to describe the
local absorption via a non-local susceptibility. The analogy with
ultra-fast time-resolved spectroscopies \cite{Shah}, that have
demonstrated the importance of phase coherence in the
quantum-mechanical time evolution of photoexcited carriers
\cite{Baumberg}, suggests that similar effects may occur in the space
domain.

To investigate the response of semiconductor nanostructures under
these conditions, we have recently proposed \cite{Mauritz99} a
theoretical approach based on a microscopic description of
electronic quantum states and their Coulomb interaction. 
Our approach
is intended to treat very high resolution probes, which might be
capable to reveal Coulomb induced coherence effects; therefore, we
consider an inhomogeneous EM-field distribution with a spatial
extension of the order of the Bohr radius of the material \cite{PerITestoni}. 
In this
paper we describe in detail our theoretical approach and present
absorption spectra calculated in the linear response regime
for a set of semiconductor quantum
wires (QWR) with realistic geometry and composition, focusing on T-shaped
structures as those obtained by the cleaved-edge overgrowth technique. 
We find that new features in the space-resolved optical spectra arise, 
particularly in coupled nanostructures, owing to interference effects 
in the interacting electron-hole wavefunction, and conclude that 
Coulomb effects on the local spectra must be taken into account for a
correct assignment of the experimental features. 

In Sec.\ \ref{sec:nls}, we derive the microscopic expression of the non-local 
susceptibility, including Coulomb interaction between electrons and holes, 
which is valid both in the linear and non-linear regimes. In Sec.\ \ref{sec:las}
we show how a proper definition of local absorption can be introduced 
in the case of spatially inhomogeneous EM field, which, however, 
depends on the shape of the EM-field distribution.
In Sec.\ \ref{sec:nr} we focus on the linear regime and
we apply our scheme to single and coupled wire structures, 
studying, in particular, the effects of non-locality and Coulomb-interactions 
on local spectra. 

\section{The non-local susceptibility}
\label{sec:nls}

In this section we derive a microscopic expression (i.e., based on
microscopic electron and hole wavefunctions), of the non-local optical
susceptibility $\bbox{\chi}$; this will be obtained through a
comparison between the macroscopic and the microscopic expressions for
the optical polarization of the system. The knowledge of $\bbox{\chi}$
allows to calculate the absorbed power defined in Sec.\ \ref{sec:las}.

The macroscopic polarization ${\bf P}({\bf r},t)$ induced by an
electromagnetic field ${\bf E}({\bf r},t)$ is in general given by
\begin{equation}\label{macro:pol-1}
{\bf P}({\bf r},t)=
\int d{\bf r'} \int dt'
\bbox{\chi}({\bf r},{\bf r'};t,t')
\cdot{\bf E}({\bf r'},t') \ ,
\end{equation}
where $\bbox{\chi}({\bf r},{\bf r'};t,t')$ is the non-local (both in
space and time) susceptibility tensor. When the time dependence of
$\bbox{\chi}({\bf r},{\bf r'};t,t')$ is through $t-t'$ only
(stationary regime), the above equation can be transformed into a
local equation in the frequency ($\omega$) domain, i.e.,
\begin{equation}\label{macro:pol-2}
{\bf P}({\bf r},\omega)=
\int \bbox{\chi}({\bf r},{\bf r'},\omega)
\cdot{\bf E}({\bf r'},\omega)\,d{\bf r'} ,
\end{equation}
where ${\bf E}({\bf r},\omega)$ and ${\bf P}({\bf r},\omega)$ are the
Fourier transforms of the time-dependent electric field and optical
polarization in Eq.~(\ref{macro:pol-1}).

In the usual case of a homogeneous EM-field distribution the
non-locality of $\bbox{\chi}$ is neglected, and $\bbox{\chi}
\propto \delta ({\bf
r}-{\bf r}')$ in (\ref{macro:pol-2}). In contrast, in order to
describe the response of excitonic states to an EM-field with a
spatial extension which is comparable to the Bohr radius, the
non-local character of $\bbox{\chi}$ in Eq.\ (\ref{macro:pol-2}) must be
fully retained. Note also that, contrary to bulk states, excitonic
states in a nanostructure do not have translational invariance; hence,
$\bbox{\chi}$ depends separately on the spatial coordinates $\bf r$, $\bf 
r'$ and not on the relative coordinate alone.

From a microscopic point of view the local (i.e., space-dependent) 
polarization can be written as:
\begin{equation}\label{micro:pol-1}
{\bf P}({\bf r},t)= q \left\langle {\bf \hat{\Psi}}^\dagger({\bf r},t)
\,{\bf r}\,
{\bf \hat{\Psi}}({\bf r},t) \right\rangle,
\end{equation}
where $q$ is the electronic charge, 
$\left\langle \ldots \right\rangle$ denotes 
a proper ensemble average, 
and the field operator ${\bf \hat{\Psi}}({\bf r},t)$ in the Heisenberg 
picture describes the microscopic time evolution of the carrier system. 

Since in this paper we shall mainly focus on optical 
(i.e., electron-hole pairs) excitations, it is convenient to 
work within the so-called electron-hole picture.
This corresponds to writing the field operator 
${\bf \hat{\Psi}}({\bf r},t)$ 
as a linear combination of electron and hole single-particle states:
\begin{equation}\label{Psi}
{\bf \hat{\Psi}}({\bf r},t) = \sum_e \hat{c}^{ }_e (t) \Psi^{ }_e({\bf r}) 
+ 
\sum_h \hat{d}^\dagger_h (t) \Psi^*_h({\bf r}) \ ,
\end{equation}
where $\hat{c}_e$ and $\hat{d}_h$ denote destruction operators for an
electron in state $e$ and a hole in state $h$. Here, $e$ and $h$ are
appropriate sets of quantum numbers labelling the conduction and
valence states involved in the optical transition, which are described
by the single-particle wavefunctions $\Psi_{e/h} ({\bf r})$ and energy
levels $\epsilon_{e/h}$ .

By inserting the above electron-hole expansion into Eq.~(\ref{micro:pol-1}),
and neglecting intraband contributions (absent for the case of optical 
excitations), the local polarization can be rewritten as:
\begin{equation}\label{micro:pol-2}
{\bf P}({\bf r},t) = \sum_{eh} \left[p_{eh}(t) 
{\bf M}^*_{eh}({\bf r}) + \mbox{c.c.}\right]\ , 
\end{equation}
where
\begin{equation}\label{M}
{\bf M}_{eh}({\bf r}) = q\,\Psi^*_e({\bf r})\,{\bf r}\,\Psi^*_h({\bf r}) 
\end{equation}
is the local (i.e., space-dependent) dipole matrix element, and 
$p_{eh}(t) = \langle \hat{d}_h \hat{c}_e \rangle$
are non-diagonal (i.e., interband) elements of the single-particle density 
matrix, also referred to as interband polarizations.

Within the mean-field Hartree-Fock approximation,
the time evolution of the above interband polarizations $p_{eh}(t)$ is 
described by the so-called semiconductor Bloch equations
(SBEs) \cite{H-K,Kuhn}:
\begin{mathletters}
\label{SBE}
\begin{eqnarray}
\frac{\partial}{\partial t} p^{ }_{eh} &=& 
\frac{1}{i\hbar} \sum_{e'h'}
( {\cal E}^{ }_{ee'} \delta_{hh'}
+ {\cal E}^{ }_{hh'} \delta_{ee'})
p^{ }_{e'h'} \nonumber\\
&&+ \frac{1}{i\hbar} {\cal U}^{ }_{eh} 
(1 - f_{e} -f_{h}) +
\frac{\partial p^{ }_{eh}}{\partial t}|_{coll} , \label{SBEp} \\
\frac{\partial}{\partial t} f_{e} &=& \frac{1}{i\hbar} \sum_{h'}
({\cal U}^{ }_{eh'} p^*_{eh'} -
{\cal U}^*_{eh'} p^{ }_{eh'}) +
\frac{\partial f_{e}}{\partial t}\bigl|_{coll} , \label{SBEfe} \\
\frac{\partial}{\partial t} 
f_{h} &=& \frac{1}{i\hbar} \sum_{e'}
({\cal U}^{ }_{e'h} p^*_{e'h} -
{\cal U}^*_{e'h} p^{ }_{e'h}) +
\frac{\partial f_{h}}{\partial t}\bigl|_{coll} , \label{SBEfh} 
\end{eqnarray}
\end{mathletters}
where $f_e = \langle \hat{c}^\dagger_e \hat{c}^{ }_e \rangle$ and $f_h
= \langle \hat{d}^\dagger_h \hat{d}^{ }_h \rangle$ denote electron and
hole distribution functions, i.e., diagonal density-matrix elements.
Here,
\begin{eqnarray}
\label{renorm2}
{\cal E}^{ }_{ee'} &=& 
\epsilon^{ }_{e} \delta_{ee'} -
\sum_{e''} V_{ee''e'e''} f_{e''} , \\
{\cal E}^{ }_{hh'} &=& 
\epsilon^{ }_{h} \delta_{hh'} -
\sum_{h''} V_{hh''h'h''} f_{h''} ,
\end{eqnarray}
and
\begin{equation}
\label{renorm1}
{\cal U}^{ }_{eh} = U_{eh} - 
\sum_{e'h'}
V_{eh'he'} p^{ }_{e'h'} 
\end{equation}
are, respectively, the electron/hole and Rabi energies renormalized by
the Coulomb interaction \cite{H-K,rossi96,rossi98,nota-prl}, and
\begin{equation}
V_{ijkl} = \int d{\bf r} \int d{\bf r'} 
\Psi_i^*({\bf r}) \Psi_j^*({\bf r'}) V({\bf r}-{\bf r'})
\Psi_k({\bf r'}) \Psi_l({\bf r}) 
\end{equation}
are the matrix elements of the three-dimensional Coulomb interaction
$V({\bf r}-{\bf r'})$ within the single-particle electron-hole
representation.  The last (collision) term in Eqs.\ (\ref{SBE})
accounts for incoherent (i.e., scattering and diffusion) processes
\cite{elsasser}.

In the usual 
case of a homogeneous (i.e., space-independent) optical excitation 
${\bf E}^\circ$ the Rabi energy $U_{eh}$ within the dipole approximation is
given by 
\begin{equation}\label{U1}
U_{eh}(t) = -{\bf M}^{\circ}_{eh} \cdot {\bf E}^\circ(t) ,
\end{equation}
where
\begin{equation}\label{M0}
{\bf M}^\circ_{eh} = \int {\bf M}_{eh}({\bf r}) d{\bf r}
\end{equation}
is the total dipole matrix element.
In contrast, for the case of a local optical excitation ${\bf E}({\bf r})$ 
---the one considered in this paper--- the electromagnetic field cannot be 
factorized as in Eq.~(\ref{U1}); 
If, however, the space variation of the field is still
negligible on the atomic scale, the Rabi energy for a local 
excitation is given by \cite{Stahl}:
\begin{equation}\label{U2}
U_{eh}(t) = -\int {\bf M}_{eh}({\bf r}) \cdot {\bf E}({\bf r},t) d{\bf r} .
\end{equation}

Let us now focus on the stationary solutions of the SBEs (\ref{SBE}).
They can be easily found in the so-called quasi-equilibrium regime, i.e., 
by assuming equilibrium distribution functions $f_e$, $f_h$ which, 
therefore, do not depend on time; 
let us define the index $l=(e,h)$ and the matrices
\begin{mathletters}
\begin{eqnarray}
T_{ll'} & = & {\cal E}_{ee'}\delta_{hh'}+{\cal E}_{hh'}\delta_{ee'} , \\
W_{ll'} & = & V_{eh'he'} (1-f_e-f_h) , \\
S_{ll'} & = & T_{ll'} - W_{ll'} .
\end{eqnarray}
\end{mathletters}
Then, Eq.\ (\ref{SBEp}) can be rewritten as 
\begin{equation}
\frac{\partial p_l(t)}{\partial t} = \frac{1}{i\hbar} \sum_{l'}
S_{ll'} p_{l'}(t) + \frac{1}{i\hbar}  \bar{U}_l(t) ,
\label{SBEsp}
\end{equation}
where $\bar{U}_l(t) = U_{eh}(t) (1-f_e-f_h)$. 

Let us suppose that $c^\lambda_l$ and $\Sigma^\lambda$ are the
eigenvectors and eigenvalues, respectively, of the matrix $S_{ll'}$;
note that, in general, $\Sigma^\lambda$ is complex.  The eigenvector
components $c^\lambda_{eh}$ are the matrix elements of the
unitary transformation connecting our original non-interacting basis
$|eh\rangle$ with the excitonic basis $|\lambda\rangle$,
$ c^\lambda_{eh} = \langle eh \vert \lambda \rangle $.
By applying this unitary transformation,
we can rewrite Eq.\ (\ref{SBEsp}) in the excitonic basis,
\begin{equation}
\frac{\partial p^\lambda(t)}{\partial t} = \frac{1}{i\hbar} \Sigma^\lambda
p^\lambda (t) + \frac{1}{i\hbar}  \bar{U}^\lambda (t) ,
\label{SBEex}
\end{equation}
where
\begin{eqnarray}
p^\lambda (t) & = & \sum_l c_l^{\lambda *} p_l(t) , \\
\bar{U}^\lambda(t) & = & \sum_l c_l^{\lambda *} \bar{U}_l (t) .
\end{eqnarray}
If we Fourier transform Eq.\ (\ref{SBEex}) we find
\begin{equation}
p^\lambda (\omega) = - \frac{\bar{U}^\lambda (\omega)}{\Sigma^\lambda-
\hbar\omega} ,
\label{ss}
\end{equation}
$p^\lambda (\omega)$ and $\bar{U}^\lambda (\omega)$ being the Fourier 
transforms of $p^\lambda (t)$ and $\bar{U}^\lambda (t)$, respectively.

Let us consider again the local polarization field ${\bf P}({\bf r},t)$ 
in (\ref{micro:pol-2}), which in our excitonic picture $\lambda$, can be 
rewritten as
\begin{eqnarray}
{\bf P}({\bf r},t) & = &
\sum_{\lambda} \left[{\bf M}^{\lambda *}({\bf r}) p^\lambda(t) 
+ \mbox{c.c.}\right] \nonumber\\
& = & \sum_{\lambda} \int^{+\infty}_{-\infty} 
\left[{\bf M}^{\lambda *}({\bf r}) p^\lambda(\omega) +
      {\bf M}^{\lambda}({\bf r}) p^{\lambda *}(-\omega)\right]
e^{-i\omega t} d\omega , \label{micro:pol-3}
\end{eqnarray}
with the definition
${\bf M}^\lambda({\bf r}) = \sum_{l} c^{\lambda *}_{l} {\bf M}_{l}({\bf 
r}) $.
By inserting the stationary solution (\ref{ss}), 
the dipole 
matrix element ${\bf M}^\lambda({\bf r})$, and $\bar{U}^\lambda(\omega)$
we obtain
\begin{eqnarray}
{\bf P}({\bf r},\omega) = \int d{\bf r}' \sum_{\lambda,eh,e'h'} & &
c^{\lambda}_{eh} {\bf M}^*_{eh}({\bf r}) \times 
c^{\lambda *}_{e'h'} {\bf M}_{e'h'}({\bf r}') 
\left(1-f_{e^\prime}-f_{h^\prime}\right) \times \nonumber \\
 & & \left[\frac{1}{\Sigma^\lambda -\hbar\omega}+
      \frac{1}{\Sigma^{\lambda *} +\hbar\omega}\right]
\,\cdot\, {\bf E}({\bf r}',\omega),\label{micropol}
\end{eqnarray}
${\bf P}({\bf r},\omega)$ being the Fourier transform 
of ${\bf P}({\bf r},t)$.
The above microscopic result has exactly the form of the macroscopic
polarization in Eq.~(\ref{macro:pol-1}), thus providing the desired
microscopic expression for the non-local optical susceptibility tensor
$\bbox{\chi}$. If we neglect the non-resonant term in (\ref{micropol})
(rotating wave approximation) we obtain
\begin{equation}\label{chi1}
\bbox{\chi}({\bf r},{\bf r'},\omega) =
\sum_{\lambda,eh,e'h'}
{
c^{\lambda}_{eh} {\bf M}^*_{eh}({\bf r}) \times 
c^{\lambda *}_{e'h'} {\bf M}_{e'h'}({\bf r}') 
\left(1-f_{e^\prime}-f_{h^\prime}\right)
\over
\Sigma^\lambda -\hbar\omega
} \ .
\end{equation}

The above general expression describes the response of the system at the
microscopic level, provided that the the single-particle wavefunctions
entering the local dipole matrix elements ${\bf M}_{eh}({\bf r})$ are
available. For the description of the response to a local probe with
the extension comparable to the Bohr radius in a typical semiconductor,
like GaAs, it is sufficient to describe the electron and heavy-hole states
within the envelope function approximation, including 
fluctuations of the wavefunctions at the atomic scale only through bulk
parameters. Assuming isotropic electron and heavy-hole energy dispersion, we
write, as usual \cite{holes}, 
$\Psi_e({\bf r}) = u_c({\bf r}) \psi_e({\bf r})$ and
$\Psi_h({\bf r}) = u_v({\bf r}) \psi_h({\bf r})$,
where $\psi_{e/h}({\bf r})$ are electron/hole envelope functions, and
$u_{c/v}({\bf r})$ are the atomic bulk wavefunctions at the
conduction/valence edge. In this paper we consider only EM fields with
a frequency corresponding to interband transition.
Therefore, interpreting the space variables
$\bf r$, $\bf r'$ in Eq.\ (\ref{micropol}) as coarse grained 
at the atomic scale, we can write
\begin{equation}\label{Mef}
{\bf M}_{eh}({\bf r}) = {\bf M}_b \psi^*_e({\bf r}) \psi^*_h({\bf r}) \ ,
\end{equation}
where ${\bf M}_b = \Omega_c^{-1}\int_{\Omega_c} u_c({\bf r}) 
{\bf r} u_v({\bf
r}) d{\bf r}$ is the bulk dipole matrix element, with $\Omega_c$ the
volume of the unit cell. Within
such approximation scheme, the susceptibility tensor $\bbox{\chi}$ in
(\ref{chi1}) becomes diagonal, with identical elements given by
\begin{equation}\label{chi2}
\chi({\bf r},{\bf r'},\omega)=\vert M_b \vert^2 
\sum_{\lambda,eh,e'h'}c^\lambda_{eh} c^{\lambda *}_{e'h'}
(1-f_{e'}-f_{h'})
\frac{\psi_e({\bf r})\psi_h({\bf r})
\psi^{*}_{e'}({\bf r'})\psi^{*}_{h'}({\bf r'})}{
\Sigma^\lambda-\hbar\omega}.
\end{equation}

\section{Local-absorption spectrum} 
\label{sec:las}

Given the susceptibility function in (\ref{chi2}), the total absorbed power
in a generic semiconductor structure can be evaluated according to:
\begin{equation}\label{alpha1}
W(\omega) \propto 
\int d{\bf r}\,\int d{\bf r'}\, 
\Im\left[
E({\bf r},\omega) \chi({\bf r},{\bf r'},\omega)
E({\bf r'},\omega) 
\right]
\ .
\end{equation}
In the usual definition of the absorption coefficient 
within the dipole approximation the non-locality of $\chi$
is neglected: $\chi({\bf r,r}') \propto \delta({\bf r-r}')$.
When non-locality is taken into account, it is no longer possible to
define an absorption coefficient that locally relates the absorbed
power density with the light intensity. 

However, considering a light 
field with a given profile $\xi$ centered around the beam position 
${\bf R}$, $E({\bf r},\omega)=E(\omega)\xi({\bf r}-{\bf R})$, 
we may define a local absorption that is a function of the
beam position, and relates the {\em total}
absorbed power to the power of a {\em local} excitation 
(illumination mode):
\begin{equation}\label{alpha2}
\alpha_\xi({\bf R},\omega)
\propto
\int 
\Im\left[  
\chi({\bf r},{\bf r'},\omega)\right] 
\xi({\bf r}-{\bf R})\xi({\bf r'}-{\bf R})\,d{\bf r}\,d{\bf r'} \ .
\end{equation}

This expression is in principle not limited to low-photoexcitation
intensities; via $f_e, f_h$ appearing in Eq.\ (\ref{chi2}) it provides
a general description of linear as well as non-linear local response,
i.e., from excitonic absorption to the gain regime. On the other hand,
in the linear-response regime $1-f_e-f_h\simeq1$ and the quantity
$\Psi^\lambda({\bf r_e},{\bf r_h})= \sum_{eh}
c^\lambda_{eh}\psi_e({\bf r_e})\psi_h({\bf r_h})$ can be identified
with the exciton wavefunction; in this case the explicit form of the
local-absorption coefficient (\ref{alpha2}) can be written as
\begin{equation}\label{alpha3}
\alpha_\xi({\bf R},\omega) = \Im
\left[\sum_\lambda \frac{\alpha_\xi^\lambda({\bf R},\omega)}
{\Sigma^\lambda-\hbar\omega}\right]\ ,
\end{equation}
where
\begin{equation}\label{exc_contrib} 
\alpha^\lambda_\xi({\bf R},\omega) 
\propto
\left\vert \int \Psi^\lambda({\bf r},{\bf r})\xi({\bf r}-{\bf R})
\,d{\bf r}\, \right\vert^2.
\end{equation}
The effects of spatial coherence
of quantum states are easily understood in the linear regime on the
basis of Eq.\ (\ref{exc_contrib}). For a spatially homogeneous
EM-field, the absorption spectrum probes the average of
$\Psi^{\lambda}$ over the whole space (global spectrum).  In the
opposite limit of an infinitely narrow probe beam,
$\alpha^\lambda_\xi({\bf R}, \omega$) maps $\vert
\Psi^{\lambda}\vert^2$; the local absorption is non-zero at any point
where the exciton wavefunction gives a finite contribution.  It is
therefore clear that ``forbidden'' excitonic transitions, not present
in the global spectrum, may appear in the local one.  In the
intermediate regime of a narrow but finite probe, it is possible that
a cancellation of the contributions from $\Psi^{\lambda}$ at different
points in space takes place, leading to a non trivial localization of
the absorption. The result will then be quite sensitive to the
extension of the light beam.

\section{Numerical results}
\label{sec:nr}

The theoretical formulation of Secs.\ \ref{sec:nls} and
\ref{sec:las} is valid for semiconductors of arbitrary
dimensionality. To illustrate the effects of non-locality and Coulomb
interaction on the local absorption spectrum, we now consider
quasi-one-dimensional (1D) nanostructures (quantum wires), subject to
a local EM excitation propagating parallel to the free axis of the
structure, $z$. For simplicity, we describe the narrow light beam by a
gaussian EM-field profile, $\xi({\bf r}) = \exp[-(x^2+y^2)/2\sigma^2]$
\cite{EM-field}.  The explicit expressions for quasi-1D systems are
derived in the Appendix.

As a prototype system, we have chosen to investigate systems composed 
of GaAs/AlAs T-shaped QWRs, which rank among the best available samples 
from the point of view of optical properties, and allow for a strong 
quantum confinement \cite{rossi97,params}.

In Fig.\ \ref{Fig:1T}(a) we show the ground-state effective wavefunction
[Eq.\ (\ref{EffWfc})] for a single QWR, including 
the electron-hole interaction; the exciton is strongly localized at the 
intersection of the parent QW's, the localization being dominated by 
that of the hole state which has a heavier effective mass \cite{rossi97}.  
When the effect of a locally inhomogeneous EM
field with a gaussian shape ($\sigma=10$ nm) is simulated
[Fig.\ \ref{Fig:1T}(b)], we find that the signal exibits a maximum at
the location of the exitonic wavefunction, but the details of the
shape of the wavefunction are lost as, for this particular sample,
they take place on a scale
shorter than $\sigma$.

The above situation for a single QWR can be contrasted with the
situation for two coupled QWRs. In the latter case the exitonic
states of the two QWRs are coupled by Coulomb interaction if their
mutual distance is $\sim a_B$; therefore, in this case the non-local
character of the Coulomb interaction can be exposed by a local probe
with $\sigma\sim a_B$. To exemplify this, we show in Fig.\ \ref{Fig:2T}
the effective wavefunction of (a) the ground and (b) the
first excited excitonic states for two coupled, symmetric QWRs,
including electron-hole Coulomb interaction. It should be noted that
1) the two excitonic states confined in the two QWRs are strongly
coupled by effect of the Coulomb interaction and 2) the effective
wavefunction is not positive definite but is even or odd for the
ground and first excited state, respectively; as consequence, in a
homogenous EM field only the ground state appears in the spectrum,
while the first excited state is prohibited by a selection rule
arising from the cancellation between positive and negative regions
[see Eq.\ (\ref{exc_contrib})]. This selection rule is relaxed in a
local optical spectroscopy experiment, if variations of the EM field
takes place on a scale comparable to the modulations of the effective
wavefunction.  In fact, when the center of mass of the beam does not
coincide with a symmetry point of the structure, the symmetry of the
whole system is broken; consequently, cancellations do not take place
exactly; moreover, they are a function of the position and extension of
the beam.

It should be stressed that the spatial dependence of the absorption in
the coupled QWR structure is dominated by Coulomb interaction, thus
making very high resolution local optical spectroscopy a very powerful
tool. To demonstrate this aspect, we have simulated a local optical
spectroscopy experiment by calculating the local absorption spectra
while sweeping the tip of the probe across a double-QWR structure; the
influence of inter-wire Coulomb interaction is demonstrated in Fig.\
\ref{Fig:panel}, where full calculations including electron-hole
interaction (center panels) are compared with calculations where the
correlation is switched off (left panels). For this example we have
chosen a set of asymmetric structures composed of two sligthly
different QWRs, with various distances between the stems of the
wires. In the uncorrelated spectra we can only distinguish two peaks
arising from single-particle transitions localized in either wires;
the two peaks shift in energy as a functions of the inter-wire
distance, decreasing from top to bottom, as a result of the increasing
overlap between the single-particle states localized in the two QWRs,
and are accompained by a high-energy tail which is due to the single
particle joint density of states. Note that there is no sign of
spatially indirect transitions connecting an electron and a hole
localized in different wires.

The situation is very different when
Coulomb correlation is taken into account. First, we
note that for the larger wire separation (top row) 
(i) the two main peaks, arising from a direct transitions located in either
wires, are red-shifted by the exciton binding energy, and (ii) the high-energy
continua are suppressed, as expected from previous studies of total
absorption in quasi-1D structures \cite{rossi96}. When the inter-wire 
distance is decreased, new peaks appear in
the spectra whose energy, intensity and location is strongly dependent
upon the coupling between the two wires, which increases from top to
bottom in Fig.\ \ref{Fig:panel}. These peaks result from interference 
between positive and negative regions of the effective wavefuctions,
whose square modulus is shown in the right column for comparison.
Figure \ref{Fig:spectra} compares the local spectra obtained with a
tip position located in the center of the right and left wire with the
total absorption for the same set of coupled QWRs as in 
Fig.\ \ref{Fig:panel} \cite{cut}.

\section*{Conclusions}

In summary, we have developed a general formulation of the theory
of local optical absorption in semiconductor nanostructures, taking into 
account quantum confinement of electron and hole states and the 
electron-hole Coulomb interaction. We have proved that absorption is 
strongly influenced by the spatial interference in the exciton wavefunctions, 
which depends on the profile of the light beam. When the extension of the
beam becomes comparable with the exciton Bohr radius, local spectra are 
expected to display different features with respect to integrated spectra, 
resulting from breaking of selection rules. Calculations performed for a 
set of coupled quantum wires show that the interpretation of near-field 
experiments will require a quantitative treatment of these effects as their
spatial resolution increases.
\acknowledgments

We thank C. Simserides for a careful reading of this manuscript.
This work was supported in part by INFM through PRA "PROCRY", 
and by the EC under the TMR Network "Ultrafast" and the IT-project 
"SQID". 

\appendix

\section*{Calculation of the local absorption in the linear regime for
quasi-1D systems}

For a QWR the single-particle electron and hole envelope functions,
appearing in (\ref{chi2}), can be written as
$\psi_e({\bf r}) = \phi^e_{\nu_e}(x,y)e^{i k^e_z z}$ and
$\psi_h({\bf r}) = \phi^h_{\nu_h}(x,y) e^{i k^h_z z}$, 
respectively, 
where $\nu_{e/h}$ and $k^{e/h}_z$ are subband indices and wavevectors
along the free axis. The envelope functions
$\phi^{e/h}_\nu(x,y)$ are solutions of a Schr\"odinger equation
with effective masses and band parameters appropriate for
electron/heavy-holes in the 2D confinement potential of the QWR.

In the linear regime the local absorption can be written as [Eq.\
\ref{exc_contrib}] 
\begin{equation}\label{absXY}
\alpha_\xi(X,Y,\omega)\propto
\sum_\lambda \Big\vert \int\Phi^\lambda(x,y)\xi(x-X,y-Y)\,dx\,dy
\Big\vert^2 h(\omega-\omega^\lambda).
\end{equation}
where $\omega^\lambda$ is the resonance frequency,
$h(\omega)$ describes the line broadening, and
\begin{equation}
\Phi^\lambda(x,y)\equiv\int \Psi^\lambda({\bf r},{\bf r})\,dz.
\end{equation}
We shall refer to $\Phi^{\lambda}(x,y)$ as the {\em effective exciton
wavefunction}; according to Eq.\ (\ref{absXY}), when convoluted with
the spatial distribution of the EM-field, $\xi(x-X,y-Y)$,
$\Phi^{\lambda}(x,y)$ yields the contribution of the $\lambda$-th
excitonic state to the local absorption $\alpha_\xi(X,Y,\omega)$;

Taking advantage of the translational invariance along $z$ we have
\begin{equation} \label{EffWfc}
\Phi^{\lambda}(x,y) = \sum_{\nu_e\nu_h} P^\lambda_{\nu_e \nu_h }
\phi^e_{\nu_e}(x,y)\phi^h_{\nu_h}(x,y),
\end{equation}
where we have defined 
$P^\lambda_{\nu\nu'} = \sum_{k_z} c^\lambda_{\nu k_z \nu' -k_z}$.
Note that only Fourier components of the polarization with 
$k^e_z=-k^h_z$ contribute to the absorption. 

In our calculations we use a plane-wave basis set to represent the
$(x,y)$ dependence of the single-particle wavefunctions:
\begin{equation}\label{def:sgp}
\phi^{e/h}_\nu(x,y)=\frac{1}{\sqrt{L_xL_y}} 
\sum_{n_x n_y} c^{e/h,\nu}_{n_xn_y} e^{i(k_xx+k_yy)},
\end{equation}
where $k_\alpha= 2 \pi n_\alpha/ L_\alpha$ and $\alpha=x,y$.
From Eqs.\ (\ref{EffWfc}) and (\ref{def:sgp}) we get
\begin{equation}
\Phi^\lambda(x,y) = 
\sum_{\nu_e \nu_h} P^\lambda_{\nu_e \nu_h}\sum_{n^e_x n^e_y}
c^{e,\nu_e}_{n^e_xn^e_y}
e^{i(k^e_xx+k^e_yy)}\sum_{n^h_xn^h_y}
c^{h,\nu_h}_{n^h_xn^h_y}
e^{i(k^h_xx+k^h_yy)}.
\end{equation}
Therefore, we can write
\begin{equation}
\Phi^\lambda(x,y)=\sum_{n_xn_y} C^\lambda_{n_xn_y}e^{i(k_xx+k_yy)},
\end{equation}
where the Fourier coefficients $C^\lambda_{n_xn_y}$ are given by
\begin{equation}
C^\lambda_{n_xn_y}=\sum_{\nu_e\nu_h} P^\lambda_{\nu_e\nu_h} 
\left(\sum_{n^e_x n^e_y n^h_x n^h_y}^{\mbox{\hspace{8truemm}}\prime} 
c^{e,\nu_e}_{n^e_xn^e_y}c^{h,\nu_h}_{n^h_xn^h_y}\right),
\end{equation}
and the primed summation is subjected to the conditions
$n^e_x+n^h_x=n_x, n^e_y+n^h_y=n_y$.

Finally, if $\xi$ can be factorized as
$\xi(x,y)=\xi_x(x)\xi_y(y)$, as is the case of a gaussian,
then the integral in Eq.~(\ref{absXY}) is given by
\begin{equation}
\int\Phi^\lambda(x,y)\xi(x-X,y-Y)\,dx\,dy=
\sum_{n_xn_y}C^\lambda_{n_xn_y}\hat\xi_x(k_x)\hat\xi_y(k_y)
e^{i(k_xX+k_yY)},
\end{equation}
where 
\begin{equation}
\hat\xi_\alpha(k_\alpha)=\frac{1}{2\pi}\int\xi_\alpha(\alpha)
e^{-ik_\alpha\alpha}\,d\alpha.
\end{equation}

\begin{figure}
\noindent
\unitlength1mm
\begin{picture}(130,170) 
\put(40,20){\psfig{figure=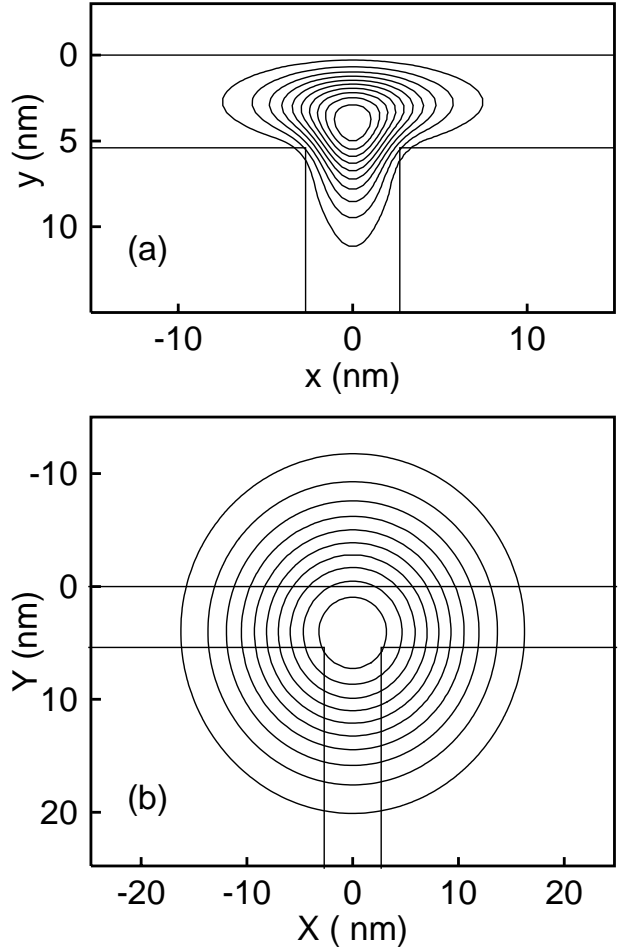,width=80mm}} 
\end{picture}
\caption{(a) Effective wavefunction [Eq.\ (\ref{EffWfc})] for the 
ground-state exciton of a single T-shaped GaAs/AlAs quantum wire 
obtained at the intersection between two quantum wells of width 5.4 nm. 
The electron-hole interaction is taken into account (12 subbands included 
in the calculation of the polarization). Only the real part is plotted; 
the imaginary part is negligible.  
(b) Contribution of the same ground-state exciton to the local absorption,
$\alpha_\xi(X,Y,\omega^\lambda)$ [see Eq.\ (\ref{absXY})], calculated for 
an EM-field$, \xi$, with gaussian distribution and $\sigma=10$ nm. 
The T-wire confinement profile is shown as a reference in 
both panels.}
\label{Fig:1T}
\end{figure}

\begin{figure}
\noindent
\unitlength1mm
\begin{picture}(130,170) 
\put(40,20){\psfig{figure=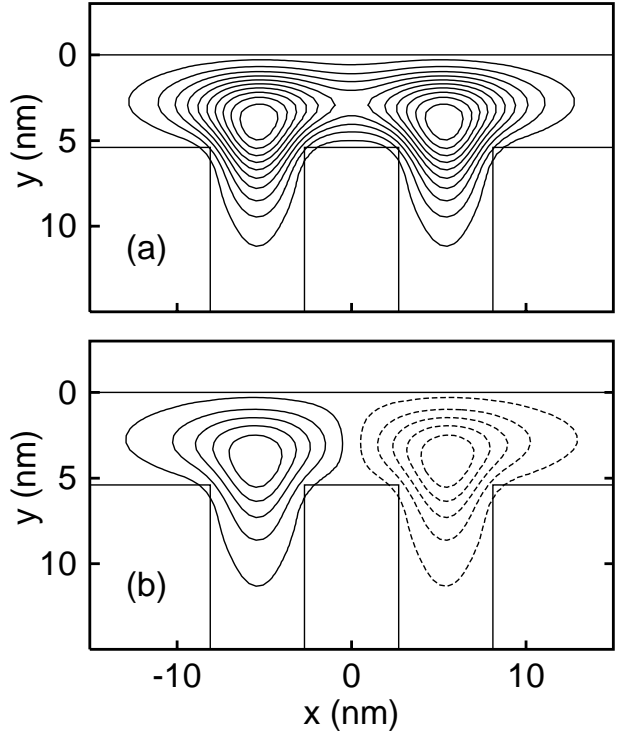,width=80mm}} 
\end{picture}
\caption{ Effective wavefunction [Eq.\ (\ref{EffWfc})] for (a)  the ground
state and (b) the first excited state in structure composed of two
symmetric T-shaped quantum wires, each obtained at the intersection of 
two GaAs/AlAs wells. The real part is plotted in (a), while 
the imaginary part is negligible; the opposite applies to (b). Both the 
parent QW's and the barrier between 
the vertical stems are 5.4 nm wide. Coulomb interaction is taken into
account (2 subbands included in the calculation of the polarization).
The confinement profile is shown in both panels for reference.}
\label{Fig:2T}
\end{figure}

\begin{figure}
\noindent
\unitlength1mm
\begin{picture}(130,170) 
\put(40,20){\psfig{figure=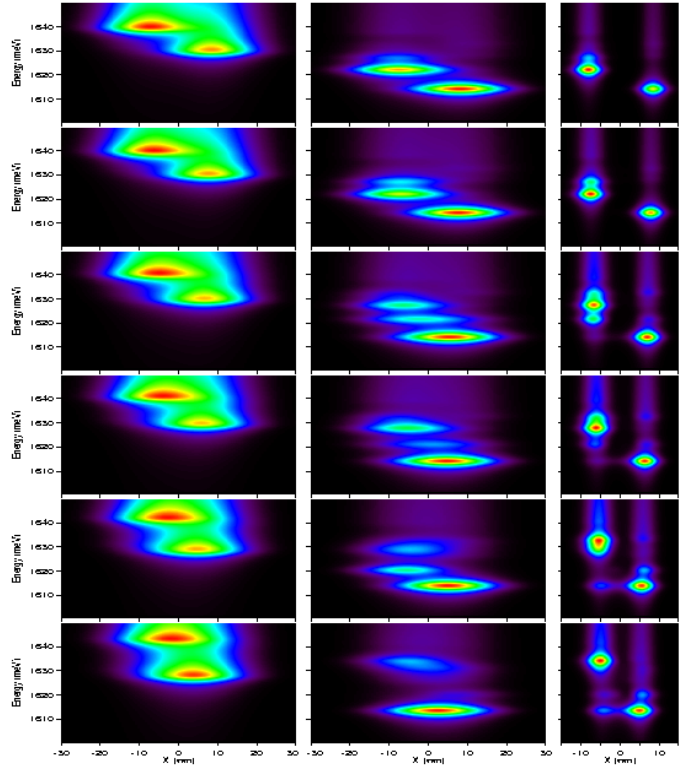,width=100mm}} 
\end{picture}
\caption{Calculated local absorption $\alpha(X,Y,\hbar\omega)$
as a function of photon energy and beam position for a set of 
asymmetric T-shaped quantum wires whose coupling is decreasing
from the top to the bottom row.
Left column: the spectra are calculated in the 
single-particle approximation. Central column: full calculation
including electron-hole Coulomb interaction.
Right column: square modulus of the effective wavefunction.
Two subbands were included in the calculations.
The three panels in each row refer to the same structure,
obtained at the intersection between an horizontal QW 
(5.4 nm wide), and two vertical QWs (the left QW is 5.4 nm wide
and the right QW is 6.0 nm wide). The vertical QWs are separated 
by a distance $d$ equal to (from top to bottom) 10.8 nm, 9.6 nm, 8.0 nm, 
6.8 nm, 5.4 nm, and 4.4 nm. 
The tip position $X$ is swept across the structure along a line positioned 
in the middle of the horizontal QW; the EM-field distribution is
gaussian, with $\sigma=10$ nm, and a broadening $\Gamma
= 2$ meV is included in the calculation.}
\label{Fig:panel}
\end{figure}

\begin{figure}
\noindent
\unitlength1mm
\begin{picture}(130,170) 
\put(40,20){\psfig{figure=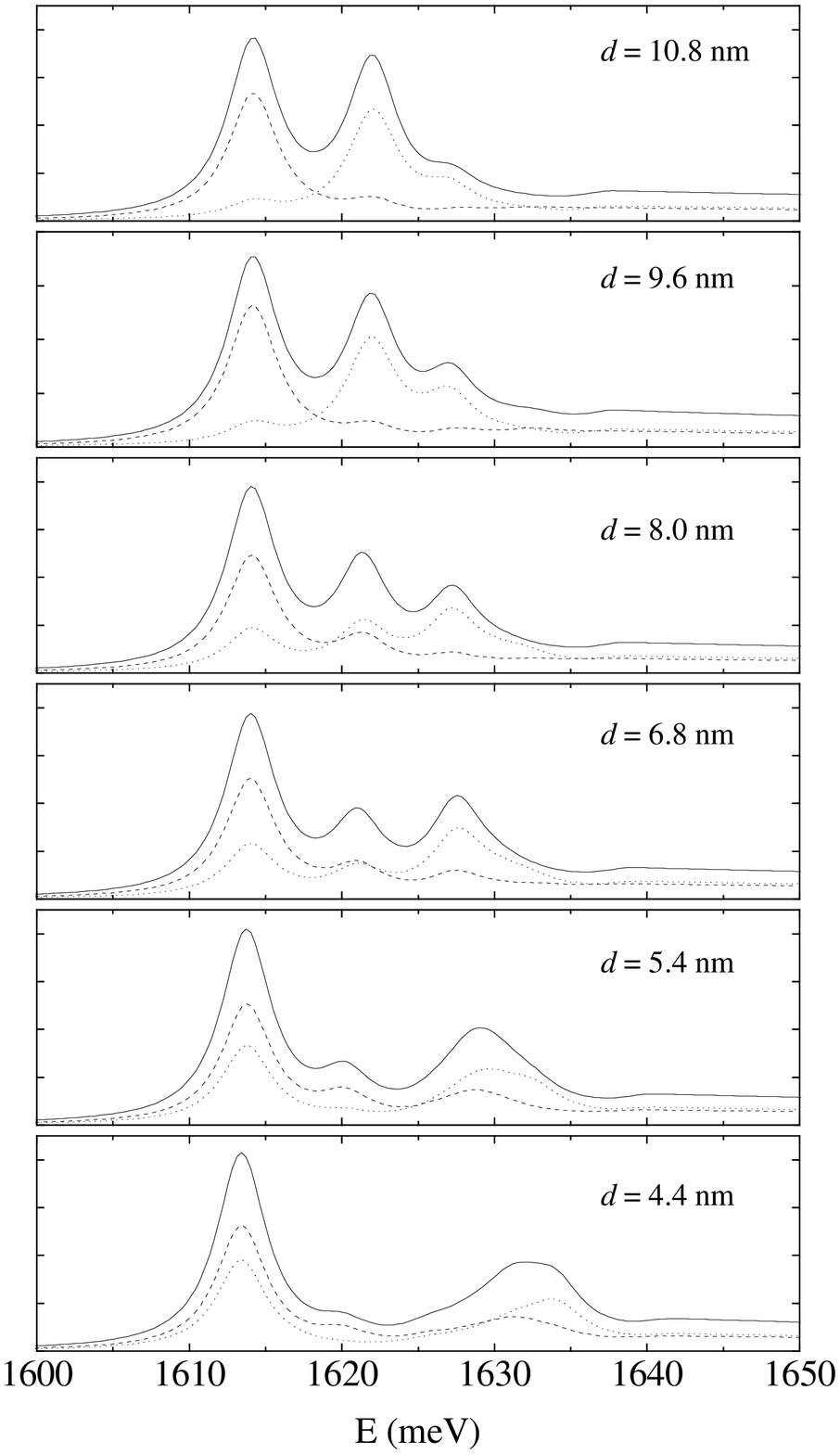,width=80mm}} 
\end{picture}
\caption{Global (solid line) and local absorption spectra
(including electron-hole correlation), calculated with the beam
centered on the right wire (dashed line) and on the left wire (dotted
line) for the same set of nanostructures as in Fig.\ \ref{Fig:panel} 
(same order from top to bottom). Here $\sigma=10$ nm and an artificial 
inhomogeneous broadening ($\Gamma = 2$ meV) is included.}
\label{Fig:spectra}
\end{figure}




\begin{references}
\bibitem{nota-prl} In Eq.\ (4) of Ref.\ \onlinecite{Mauritz99} only the
diagonal renormalized Rabi energies were included. This corresponds to
neglect intraband polarizations which are expected to be very
small. Obiuosly, off-diagonal Rabi energies are exactly zero in the
linear regime.
\bibitem{EM-field}
The actual profile of the EM field generated by a NSOM tip
is more complex, and a research topic in itself.  
However, our simple choice for $\xi$ is sufficient to highlight
the effects of nonlocality. Any field distribution
can be included in our treatment through Eq.\ (\ref{exc_contrib}), while 
Eq.\ (\ref{chi2}) is independent of the external field.
\bibitem{params} We use homogeneous and
isotropic effective masses $m^e=0.067m_0$ and $m^h=0.34m_0$;
the GaAs/AlAs band
offsets are $V^e=1.033$ eV and $V^h=0.558$ eV.
\bibitem{cut} For clarity, in the present calculations we have 
neglected higher-energy states from the wider QW. We have checked 
that these do not affect the local spectra
in Fig.\ \ref{Fig:spectra}, but would only contribute to the high-energy part
of the global spectrum in the same figure.
\end{references}
\end{document}